\documentclass[aps,prl,twocolumn,amsmath,amssymb,superscriptaddress]{revtex4-1}
\usepackage{graphicx}  % figures
\usepackage{hyperref}  % hyperlinks (and PDF comments as options)
\usepackage{amsmath}   % for fancy equations etc.

\begin{document}

\title{Temperature-dependent interplay of Dzyaloshinskii-Moriya interaction and single-ion anisotropy in multiferroic BiFeO$_3$}

\author{Jaehong Jeong}
\email{hoho4@snu.ac.kr}
\affiliation{Center for Correlated Electron Systems, Institute for Basic Science (IBS), Seoul 151-747, Korea}
\affiliation{Department of Physics and Astronomy, Seoul National University, Seoul 151-747, Korea}

\author{Manh Duc Le}
\affiliation{Center for Correlated Electron Systems, Institute for Basic Science (IBS), Seoul 151-747, Korea}
\affiliation{Department of Physics and Astronomy, Seoul National University, Seoul 151-747, Korea}

\author{P. Bourges}
\affiliation{Laboratoire L\'{e}on Brillouin, CEA-CNRS, CEA-Saclay, F-91191 Gif-sur-Yvette, France}

\author{S. Petit}
\affiliation{Laboratoire L\'{e}on Brillouin, CEA-CNRS, CEA-Saclay, F-91191 Gif-sur-Yvette, France}

\author{S. Furukawa}
\affiliation{Department of Physics, University of Tokyo, 7-3-1 Hongo, Bunkyo-ku, Tokyo 113-0033, Japan}

\author{Shin-Ae Kim}
\affiliation{Neutron Science Division, Korea Atomic Energy Research Institute, Daejeon 305-353, Korea}

\author{Seongsu Lee}
\affiliation{Neutron Science Division, Korea Atomic Energy Research Institute, Daejeon 305-353, Korea}

\author{S-W. Cheong}
\affiliation{Rutgers Center for Emergent Materials and Department of Physics and Astronomy, Rutgers University, Piscataway New Jersey 08854, USA}

\author{Je-Geun Park}
\email{jgpark10@snu.ac.kr}
\affiliation{Center for Correlated Electron Systems, Institute for Basic Science (IBS), Seoul 151-747, Korea}
\affiliation{Department of Physics and Astronomy, Seoul National University, Seoul 151-747, Korea}

\date{\today}

\begin{abstract}
Low-energy magnon excitations in multiferroic BiFeO$_3$ were measured in detail as a function of temperature around several Brillouin zone centers by inelastic neutron scattering experiments on single crystals. Unique features around 1 meV are directly associated with the interplay of the Dzyaloshinskii-Moriya interaction and a small single-ion anisotropy. The temperature dependence of these and the exchange interactions were determined by fitting the measured magnon dispersion with spin-wave calculations. The spectra best fits an easy-axis type magnetic anisotropy and the deduced exchange and anisotropy parameters enable us to determine the anharmonicity of the magnetic cycloid. We then draw a direct connection between the changes in the parameters of spin Hamiltonian with temperature and the physical properties and structural deformations of BiFeO$_3$.
\end{abstract}

\maketitle

Multiferroic compounds exhibiting phase transitions arising from the two otherwise unrelated order parameters of magnetic moment and electric polarization \cite{MF:Cheong-natmat2007} have attracted huge interest with prime examples being hexagonal manganites and BiFeO$_3$\cite{Fiebig2002,SLee2008,Ramesh,Choi2009}. Of the several multiferroic materials, BiFeO$_3$ is the only compound that exhibits multiferroicity above room temperature \cite{BFO:Catalan-advmat2009} with a ferroelectric transition at $T_\text{C} \sim$ 1100~K and an antiferromagnetic (AFM) transition at $T_\text{N} \sim$ 650~K. It is an excellent candidate for Magneto-Electric(ME) devices working at room temperature with a large electric polarization $P \sim 100~ \rm \mu C/cm^2$. Below $T_\text{N}$, an incommensurate cycloid magnetic structure is formed along the $[1,1,0]$ direction in the hexagonal notation with an extremely long period of 620~\AA ~\cite{Sosnowska1982}. In addition, the incommensurate magnetic structure is further canted out of the cycloid plane \cite{Ramazanoglu2011}, which was recently pointed out to be closely related to the magnetoelectric coupling mechanism \cite{Lee2013}.

Although it was realised early on that the microscopic interactions revealed by measurements of the spin dynamics is important to understand the complex magnetic structure and its coupling to the lattice~\cite{Sosnowska1995}, magnon excitations in BiFeO$_3$ have only been studied recently. Initial attempts were made using the Monte-Carlo method \cite{Park2011,Kenji} and THz spectroscopy \cite{Talbayev2011}, whilst we reported the full spin wave dispersion measured by inelastic neutron scattering (INS) described by a spin Hamiltonian including two exchange interactions and the DM interaction \cite{Jeong2012}. Further INS measurements additionally determined the single-ion anisotropy (SIA) \cite{Matsuda2012}. A further detailed theory was proposed to explain the spectroscopic modes seen in THz spectroscopy and INS by a spin Hamiltonian including the DM interaction along two directions and SIA \cite{Fishman1st,Fishman2nd}.

Despite the experimental and theoretical works\cite{Park2011,Kenji,Talbayev2011,Jeong2012,Matsuda2012,Fishman1st,Fishman2nd}, the detailed features of magnon excitations at low energy has not been fully examined by experiments and thermal variations of the DM interaction and SIA, most importantly a complete discussion of magnetic easy-axis or easy-plane anisotropy, still remains unexplored. One should note that the precise determination of the temperature-dependent magnetic parameters such as the exchange and DM interaction and the type of magnetic anisotropy are crucial to a full and microscopic understanding of BiFeO$_3$. In particular, this information, if determined accurately, address directly the key questions of BiFeO$_3$. We demonstrate here that the interplay of the DM interaction and SIA is essential to understanding the magnon excitations in BiFeO$_3$ at low energy, an analysis of which provide the values of these parameters at various temperatures.

\begin{figure}[htb]
\includegraphics[width=\columnwidth,clip]{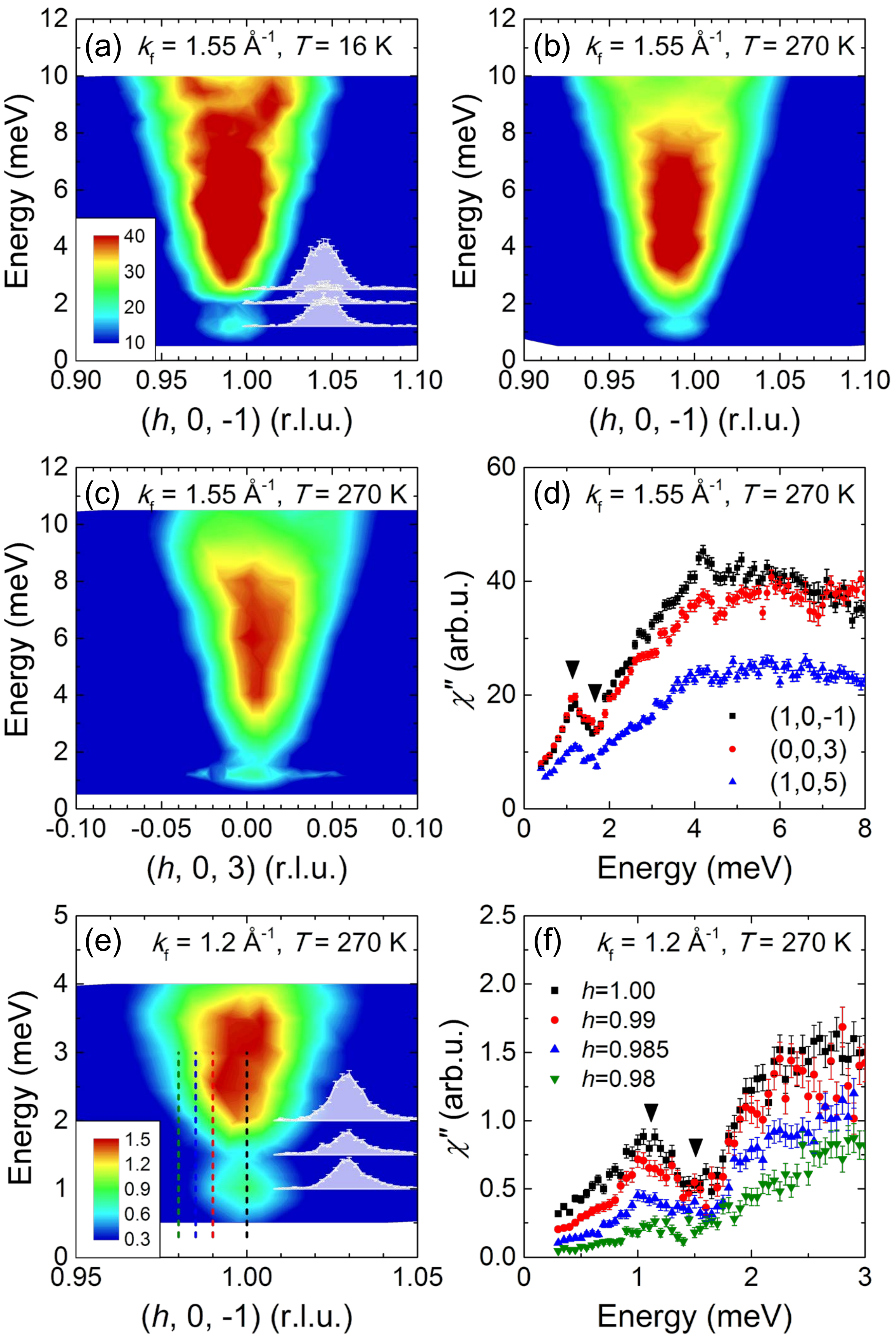}
\caption{\label{fig1}(color online) (a-c) Experimental Im$[\chi(\bold{q},\omega)]$ along the $[h,0,0]$ direction at different temperatures and Brillouin zone centers with $k_\text{f}=$ 1.55~\AA$^{-1}$. (d) Energy scans at three different $\Gamma$-points with $k_\text{f}=$ 1.55~\AA$^{-1}$. (e) Higher-resolution $q$-$E$ map at $\bold{q}=(1,0,-1)$ and $T=$ 270~K with $k_\text{f}=$ 1.2~\AA$^{-1}$. (f) $E$-scans along dashed lines in (e) with $k_\text{f}=$ 1.2~\AA$^{-1}$. In (a) and (e), $q$-scans are shown for three representative low energies.}
\end{figure}

BiFeO$_3$ has a rhombohedral structure with space group $R3c$, $a=$ 5.573 and $c=$ 13.842~\AA. An assembly of eight single crystals of total mass 1.6 g, grown by the flux method, was prepared by coaligning them within 3$^\circ$. Inelastic neutron scattering experiments were twice performed using the cold-neutron triple axis spectrometer 4F2 at Laboratoire L\'{e}on Brillouin in France. The sample was mounted with the $b$ axis normal to the horizontal scattering plane of the spectrometer, i.e., in the $a^*$-$c^*$ plane. First measurements were carried out at various temperatures and Brillouin zone centers using the $k_\text{f}$-fixed mode with $k_\text{f}=$ 1.55~\AA$^{-1}$. A Be filter was installed to remove higher neutron harmonics of the scattered beam. The contour maps of the neutron intensity as a function of energy and $\bold{Q}$ along $(h 0 -1)$ in Fig. \ref{fig1}(a-c) were obtained from successive constant-energy $q$-scans centered on $\bold{q}=(1,0,-1)$ and $(0,0,3)$ along the $[h,0,0]$ direction at $T=$ 16 and 270~K. The constant-momentum $E$-scans in Fig. \ref{fig1}(d) were collected at different Brillouin zone centers $\bold{q}=(1,0,-1)$, $(0,0,3)$ and $(1,0,5)$ with $T=$ 270~K. To examine the temperature dependence, $E$-scans at $\bold{q}=(1,0,-1)$ were measured at $T=$ 16, 50, 100, 200 and 270~K as shown in Fig. \ref{fig4}(a).
In order to examine the feature around $E$=1~meV in detail, we performed another experiment with a smaller final neutron momentum $k_\text{f}=$ 1.2~\AA$^{-1}$ for better resolution. The $q$-$E$ map along the $[h,0,0]$ direction and $E$-scans below 4~meV were collected at $\bold{q}=(1,0,-1)$ and $T=$ 270~K as shown in Fig. \ref{fig1}(e-f). In all cases, the measured intensities were corrected by the thermal population Bose factor to extract the imaginary part of the generalized magnetic susceptibility, Im$[\chi(\bold{q},\omega)]$.

Although a small modulation exists, the magnetic ground state of BiFeO$_3$ is basically G-type AFM where nearest-neighbor spins are anti-parallel, such that a typical V-shaped dispersion was expected at low energy. However, an unusual island-like shape was found in the $Q$-$E$ maps at low energy transfer, $E \sim$ 1~meV, as indicated in Fig. \ref{fig1}. This corresponds to a peak and a dip in $E$-scans at the zone center. We confirm that these unique features can be explained only by the mode coupling caused by the interplay of DM interaction and SIA.

\begin{figure*}[t]
\includegraphics[width=\textwidth,clip]{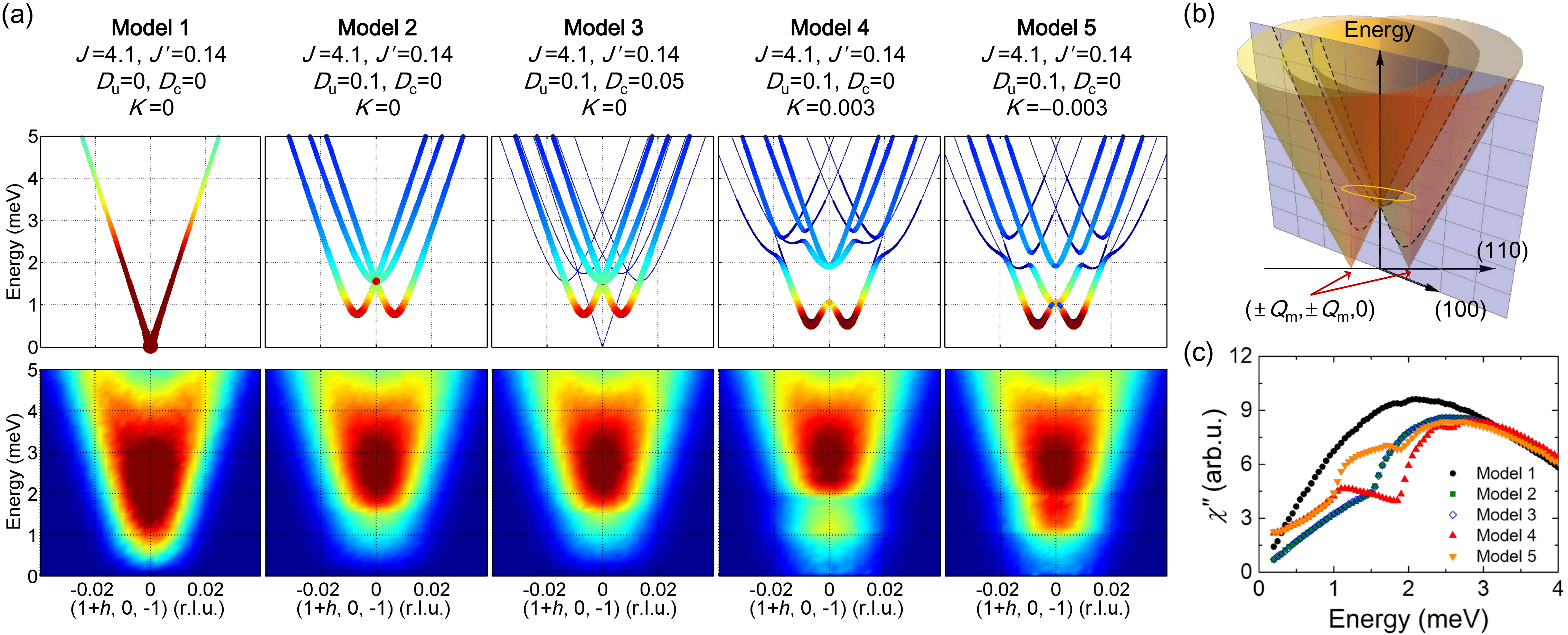}
\caption{\label{fig2}(color online) (a) Calculated magnon dispersion curves and simulated Im$[\chi(\bold{q},\omega)]$ convoluted with the instrumental resolution function along $[h,0,0]$ centered on $\bold{q}=(1,0,-1)$ for different model parameters. (b) Schematic view of the $q$-$E$ plane along $[h,0,0]$ and resolution ellipsoid (orange). (c) Simulated $E$-scan at $(1,0,-1)$ is shown for five different models.}
\end{figure*}

We previously presented a spin Hamiltonian with nearest and next-nearest neighbor exchange interactions and a DM interaction along $[1,1,0]$ in Ref. \cite{Jeong2012}. In order to explain the low-energy feature, another DM term along the $c$-axis and an SIA term were also considered, as discussed in Ref. \cite{Matsuda2012,Fishman1st,Fishman2nd}, 
\begin{equation} \label{eq:hamiltonian}
\begin{split}
\mathcal{H}  = & J \sum_{\bold{r},\boldsymbol{\alpha}} \bold{S}_{\bold{r}} \cdot \bold{S}_{\bold{r}+\boldsymbol{\alpha}} + J' \sum_{\bold{r},\boldsymbol{\beta}} \bold{S}_{\bold{r}} \cdot \bold{S}_{\bold{r}+\boldsymbol{\beta}} \\
  & -D_u \sum_{\bold{r}} \hat{\bold{u}} \cdot ( \bold{S}_{\bold{r}} \times \bold{S}_{\bold{r} + a \hat{\bold{v}}} ) \\
  & -D_c \sum_{\bold{r}} (-1)^{ {6 \bold{r} \cdot \hat{\bold{c}}}/{c} }~ \hat{\bold{c}} \cdot ( \bold{S}_{\bold{r}} \times \bold{S}_{\bold{r} + \frac{c}{2} \hat{\bold{c}}} ) \\
  & -K \sum_{\bold{r}} ( \bold{S}_{\bold{r}} \cdot \hat{\bold{c}} )^2 , 
\end{split}
\end{equation} 
where $\bold{S}_{\bold{r}}$ is the spin-5/2 operator at the position $\bold{r}$, and $\hat{\bold{u}}$, $\hat{\bold{v}}$, and $\hat{\bold{c}}$ are the unit vectors along the directions $[1,-1,0]$, $[1,1,0]$, and $[0,0,1]$, respectively.
In the first two terms for the exchange interaction, $\boldsymbol{\alpha}$ and $\boldsymbol{\beta}$ are displacement vectors for the nearest and next-nearest neighbors, respectively. The third and forth terms originate from the DM interaction induced by a distortion of the Fe-O-Fe bond. This can be effectively separated into two terms, one which acts along $\hat{\bold{v}}$ with a chiral vector $\bold{D_u} =D_u  \hat{\bold{u}}$ and another along $c$-axis with an alternate chiral vector $\bold{D_c} = (-1)^{ {6 \bold{r} \cdot \hat{\bold{c}}}/{c} } D_c \hat{\bold{c}}$. The last term in Eq. (\ref{eq:hamiltonian}) describes the SIA along the $c$-axis. Using this Hamiltonian, we calculated the full dispersion curve of spin waves by using linear spin wave theories.

We now examine five models with different parameters for $J$, $D_u$, $D_c$ and $K$. Model 1 is the simplest, with only  exchange interaction terms ($D_u=D_c=K=0$). Model 2 includes the main DM term along $[1,1,0]$ ($D_u>0$, $D_c=K=0$) as well, which gives the long period magnetic cycloid. In model 3, the additional DM term along $c$-axis is included ($D_u>0$, $D_c>0$, $K=0$), which causes a small tilting of the magnetic cycloid plane around the $c$-axis. Model 4 contains a small easy-axis SIA instead of the second DM term ($D_u>0$, $D_c=0$, $K>0$). In contrast, a small easy-plane SIA is considered ($D_u>0$, $D_c=0$, $K<0$) in model 5. The small SIA with the DM term produces a slightly anharmonic cycloid due to a modulation of the angle between adjacent spins along the cycloid axis \cite{cyb,Sosnowska2011}, which will be discussed shortly. 
The magnon dispersion relation $\omega(\bold{q})$ and dynamical structure factor $S(\bold{q},\omega)$ was calculated for each model using the Holstein-Primakoff boson operators as discussed in Ref. \cite{Jeong2012}. 

Theoretical spectra along the $[h,0,0]$ direction are given in the upper panels of Fig. \ref{fig2}(a). In model 1, three modes detected by different components of $S(\bold{q},\omega)$ are degenerate whereas these are split by the $D_u$ term in the model 2. The additional DM term, D$_c$ mixes the modes at the wave vectors $\bold{q}$ and $\bold{q}\pm \bold{Q}_m$ (with $\bold{Q}_m$=[0.0045 0.0045 0] being the incommensurate vector) and the energy dispersions are folded in a complicated way. However, the mixing amplitude vanishes asymptotically in the low-energy limit, and there is no noticeable difference in the low-energy spectrum. On the other hand, the SIA term in models 4 and 5 makes significant changes; it appreciably mixes the modes at $\bold{q}$ and $\bold{q}\pm 2\bold{Q}_m$ even at low energies, and the folded spectrum shows an energy gap at the zone center as indicated. 

For a direct comparison with the experimental data, a theoretical simulation was performed for the dynamical magnetic susceptibility Im$[\chi(\bold{q},\omega)]$ convoluted with the instrumental resolution function. Taking the resolution matrices at $E=$ 3~meV for $k_\text{f}=$ 1.55~\AA$^{-1}$ and at $E=$ 1~meV for $k_\text{f}=$ 1.2~\AA$^{-1}$ as representative, Im$[\chi(\bold{q},\omega)]$ was numerically calculated with a million $q$-points sampled in a Gaussian distribution defined by the resolution ellipsoids and summed. The resolution ellipsoid in the $q$-$E$ plane along $[h,0,0]$ are shown schematically together with the dispersion curves in Fig. \ref{fig2}(b). In the lower panels of Fig. \ref{fig2}(a), the complex mode mixing and the gap in the model 4 reproduces the unique island-like shape very well. The difference is more obvious in the simulated $E$-scan in Fig. \ref{fig2}(c). The characteristic peak at low energy appears only with coexistence of the DM interaction $D_u$ and the easy-axis SIA $K$. In model 5, we repeated the calculation for the case of easy-plane magnetic anisotropy, which noticeably fails to reproduce the island-type low energy excitation.

\begin{figure}[htb]
\includegraphics[width=\columnwidth,clip]{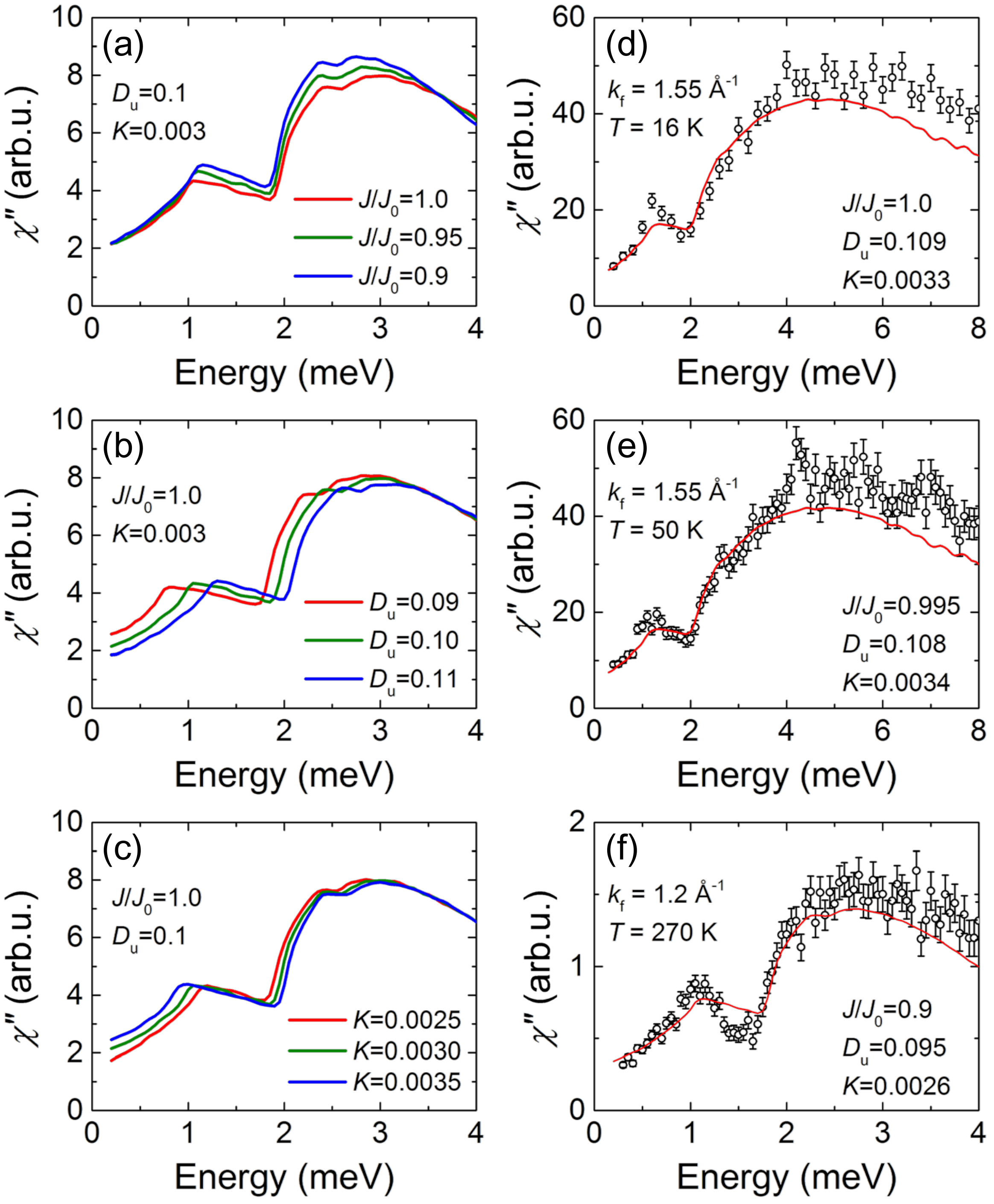}
\caption{\label{fig3}(color online) Simulation results for different (a) $J/J_0$, (b) $D_u$ and (c) $K$. Best fit results with $k_\text{f}=$ 1.55~\AA$^{-1}$ at (d) 16~K, (e) 50~K and (f) with $k_\text{f}=$ 1.2~\AA$^{-1}$ at 270~K.}
\end{figure}

In order to determine parameters of the Hamiltonian \ref{eq:hamiltonian}, we examined the effect of $J$, $D_u$ and $K$ on the simulated $E$-scan results. To simplify the problem, the ratio of $J$ to $J^\prime$ is fixed and their values at 16~K are taken from our previous letter: $J=$ 4.38, $J^\prime=$ 0.15 meV, where we used the effective spin length $S_{\mathrm{eff}}=\sqrt{\frac{5}{2}(\frac{5}{2}+1)}$ \cite{Jeong2012}: when comparing our values with that in Ref. \cite{Matsuda2012} one should convert from our use of effective $S$ value ($S_{\mathrm{eff}}$) to just $S$. The effects of varying $J/J_0$, $D_u$ and $K$ are shown in Fig. \ref{fig3}(a-c): $J_0$ is the value at 16~K. $J/J_0$ scales the intensity of the $E$-scan, $D_u$ determines the position of the peak and dip, and $K$ determines the distance between the peak and dip, i.e. the size of the gap. We determine the best fit parameters for various temperatures: at 16 K $J=$ 4.38, $J^\prime=$ 0.15, $D_u$= 0.109, and $K$= 0.0033 meV, which are shown with the experimental results in Fig. \ref{fig3}(d-f). We estimate that $D_c$ is smaller than 0.05 meV, which appears to be one order of magnitude smaller than estimated from the tilting angle of cycloid plane reported by Ref. \cite{Ramazanoglu2011}. We think that the small discrepancy between the data and our simulation, in particular the features seen around 4--8 meV in in Fig. \ref{fig3}(e), are due to the fact that in our analysis we have used the common approach of approximating the instrument resolution volume by a Gaussian ellipsoid. However, the true resolution volume is an irregular polyhedron and as such can encompass additional modes not sampled by our Gaussian approximation, leading to extra peaks in the measured spectrum which are not in the simulations, but which do not represent additional modes.

The position of the peak and dip varies with temperature as shown with the best fit curves in Fig. \ref{fig4}(a). Although the peak is almost constant, the dip energy changes with temperature. From the best fit parameters, the temperature dependence of $J\tilde{S}$, $D\tilde{S}$ and $K\tilde{S}$ was obtained as shown in Fig. \ref{fig4}(b), where $\tilde{S}$ represents the temperature-dependent normalized moment obtained from Ref. \cite{Palewicz}. Encouragingly, it agrees with the structural change that governs each interaction term in the spin Hamiltonian.
\begin{figure}[htb]
\includegraphics[width=\columnwidth,clip]{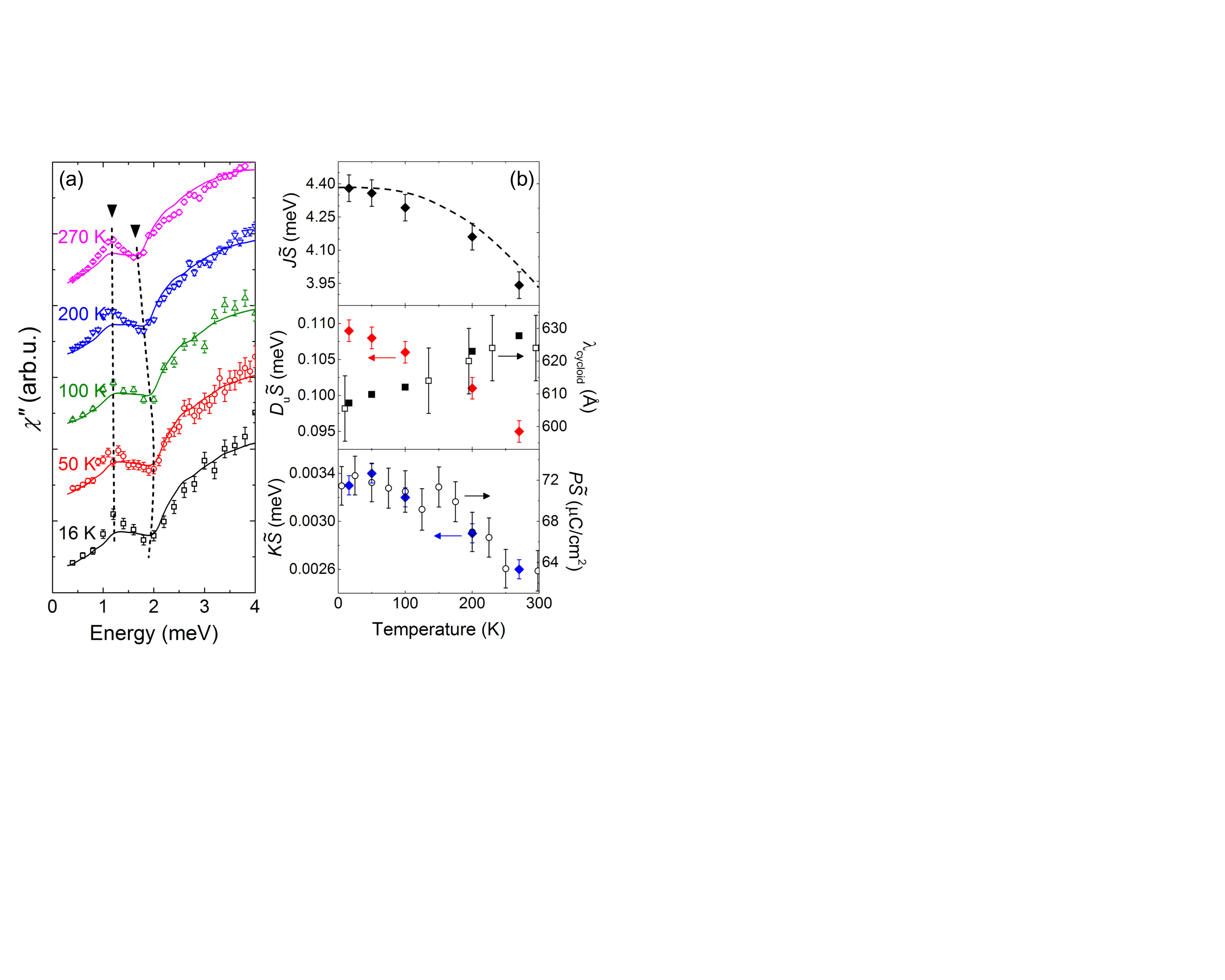}
\caption{\label{fig4}(color online) (a) Temperature dependence of Im$[\chi(\bold{q},\omega)]$ at $\bold{q}$=(1,0,-1) with best fit curves. (b) Temperature dependence of $J\tilde{S}$, $D_u \tilde{S}$, $K\tilde{S}$ and a period of the magnetic cycloid with the arrows representing typical error bars. We estimated the temperature dependence of the moment, $\tilde{S}$ (dashed line) and the electric polarization (P) using the structural data taken from Ref. \cite{Palewicz} while open symbol of $\lambda$ is taken from Ref. \cite{Sosnowska2011}.}
\end{figure}
We note that the decrease of $J\tilde{S}$, $D_u \tilde{S}$ and $K\tilde{S}$ reflects the temperature dependence of the total moment ($\tilde{S}$). According to neutron diffraction studies, the moment (dashed line in Fig. 4(b)) is reduced by 10\% from low to room temperatures, which almost entirely accounts for the observed change in $J\tilde{S}$, making $J$ almost temperature independent.

The DM interaction is proportional to the vector $\bold{r}_{\mathrm{Fe}}\times\bold{r}_{\mathrm{O}}$~\cite{zvezdin_flexoelectric_bfo}, and thus correlates with the Fe-O-Fe bond angle, which increases slightly with temperature~\cite{Palewicz}. Thus $D_u$ should decrease with temperature, as observed.
The incommensurate magnetic cycloid is nearly harmonic with a period approximately proportional to $(J-4J^\prime)/D_u$ for a small $K$. The periodicity $\lambda_\text{cycloid}$ calculated using our best fit values and the lattice parameters from Ref.~\cite{Palewicz} agrees well with the experimental results~\cite{Sosnowska2011} as shown in the middle panel of Fig.~\ref{fig4}(b). Of further interest, we could directly determine the magnetic anharmonicity value $m\simeq$ 0.58--0.64~
\footnote{The period of anharmonic cycloid can be expressed as $\lambda = 4\sqrt{m/\epsilon}~\text{K}(m)$ where $\text{K}(m)$ is the complete elliptic integral with the anharmonicity parameter $m$, and $\epsilon= 4 K/ \left( a^2 (J-4J') \right)$. For a fixed $\lambda$, $m$ can be estimated numerically.} 
using our best parameters. The cycloid anharmonicity, arising from the SIA, allows the coupling of a $\bold{q}=0$ phonon to magnons beyond the first Brillouin zone~\cite{DeSousa2008}, and is thus crucial in explaining the observations of many electromagnons in BiFeO$_3$. We note that according to our analysis the intensity of third-order satellites is about 300 times weaker than that of first-order satellites in our results, in good agreement with the diffraction result~ \cite{cyb,Sosnowska2011} although our $m$ value is found to be larger than that estimated in Ref.~\cite{Sosnowska2011}. 

The SIA $K$ is thought to be connected to the two structural distortions leading to the acentric $R3c$ space group: the ferroelectric (FE) displacement and the antiferro-distortive (AFD) rotation. It was recently pointed out by DFT calculations~\cite{spaldin} that the exact type of magnetic anisotropy is crucially dependent on the details of the local distortions of the perovskite structure and thus the size of the electric polarization  with the FE displacement favoring an easy-axis anisotropy and the AFD rotation inducing an easy-plane anisotropy. While the calculations produce a small \emph{easy-plane} anisotropy, we find experimentally that a small easy-axis anisotropy prevails. Given the precision of DFT calculations, it is unsurprising that such a small difference compared to the total energy is difficult to compute. The temperature dependence of $K$ agrees with an increase in the Fe-Bi distance, determined by neutron diffraction~\cite{Palewicz}. The increasing Fe-Bi distance both reduces the SIA and the electric polarisation which are thus correlated, as shown in Fig.~\ref{fig4}(b): the polarization is calculated by using the experimental values as in Ref.~\cite{Lee2013}. The fine sensitivity of the SIA to small structural changes may also explain the strong suppression of a magnetic domain under a modest uniaxial pressure $P\approx7$~MPa~\cite{Rama2011b}. This small pressure could affect the SIA enough to favour the other two cycloid domains, but is unlikely change the exchange interactions enough to remove the cycloid that way. Furthermore, a large SIA, whilst not realised in BiFeO$_3$, could suppress the cycloid leading to a much simpler structure like a G-type antiferromagnetism.

In summary, we confirm that the interplay of the DM interaction and easy-axis SIA is essential to explain the low-energy magnon spectra of BiFeO$_3$ measured by inelastic neutron scattering experiments. The values of $J$, $D$ and $K$ were determined at various temperatures by fitting the data, and their temperature dependence is found to be consistent with the structural changes observed by high resolution neutron diffraction \cite{Palewicz,Sosnowska2011}. Using these experimental results, we uniquely determined the exact type and temperature dependence of the magnetic anisotropy and the magnetic anharmonicity.

We thank Sanghyun Lee, Gun Sang Jeon, Heung-Sik Kim, Jaejun Yu, M. Mostovoy, and Yong-Baek Kim for useful discussion. This work was supported by the Research Center Program of IBS (Institute for Basic Science) in Korea: Grant No. EM1203. One of us (SF) was supported by JSPS KAKENHI (Grant No. 25800225) and MEXT KAKENHI (Grant No. 22103005), and SWC was supported by the DOE under Grant DE-FG02-07ER463821104484. 
 
\bibliography{BFO113-V3}

\end{document}